\documentclass[aip,apl,reprint,superscriptaddress,floatfix,amsmath,showpacs]{revtex4-1}

\usepackage{graphicx}
\usepackage{amsmath}
\usepackage{textcomp} 

\begin{document}

\title{Low-damping transmission of spin waves through YIG/Pt-based layered structures for spin-orbit-torque applications}

%
%
%
%

\author{Dmytro A. Bozhko}
\affiliation{Fachbereich Physik and Landesforschungszentrum OPTIMAS, Technische Universit\"at Kaiserslautern, 67663 Kaiserslautern, Germany}
\affiliation{Graduate School Materials Science in Mainz, Kaiserslautern 67663, Germany}

\author{Alexander A. Serga}
\email{serga@physik.uni-kl.de}
\affiliation{Fachbereich Physik and Landesforschungszentrum OPTIMAS, Technische Universit\"at Kaiserslautern, 67663 Kaiserslautern, Germany}

\author{Milan Agrawal}
\affiliation{Fachbereich Physik and Landesforschungszentrum OPTIMAS, Technische Universit\"at Kaiserslautern, 67663 Kaiserslautern, Germany}

\author{Burkard Hillebrands}
\affiliation{Fachbereich Physik and Landesforschungszentrum OPTIMAS, Technische Universit\"at Kaiserslautern, 67663 Kaiserslautern, Germany}

\author{Mikhail P. Kostylev}
\affiliation{School of Physics, M013, University of Western Australia, Crawley, WA 6009, Australia \looseness=-1}

\date{\today}

\begin{abstract}

We show that in YIG-Pt bi-layers, which are widely used in experiments on the spin transfer torque and spin Hall effects, the spin-wave amplitude significantly decreases in comparison to a single YIG film due to the excitation of microwave eddy currents in a Pt coat. By introducing a novel excitation geometry, where the Pt layer faces the ground plane of a microstrip line structure, we suppressed the excitation of the eddy currents in the Pt layer and, thus, achieved a large increase in the transmission of the Damon-Eshbach surface spin wave. At the same time, no visible influence of an external dc current applied to the Pt layer on the spin-wave amplitude in the YIG-Pt bi-layer was observed in our experiments with YIG films of micrometer thickness.

\end{abstract}

\maketitle

Magnonics addresses the transfer and processing of information using spin waves (SW) and their quanta, magnons.\cite{Chumak2015} One of the main challenges for this field is the finite lifetime of magnons even in a low-damping material such as single crystal yttrium-iron-garnet (YIG).\cite{Serga2010} A possible way to overcome this issue is the compensation of SW damping via transfer of a spin torque to a magnetic medium by a spin-polarized electron current generated in an adjacent non-magnetic metal layer with high spin-orbit interaction. The decrease in the damping of spin waves propagating in a YIG film covered with a current conducting Pt film has already been reported.\cite{Wang2011,Rezende2011} However, the initial spin-wave attenuation in such a structure was unacceptably high for practical applications ($\simeq 35$\,dB in Ref.\,\onlinecite{Wang2011}). Therefore, it is an important task to determine the origin of the excessive spin-wave damping in YIG-Pt bi-layers and find a way to reduce this harmful effect. The latter will only be possible if this damping does not correspond to the spin-pumping effect -- the process of a back transfer of a spin torque from magnons exited in a magnetic medium to a non-magnetic layer. 

Here, we show that in a Pt covered YIG film the SW amplitude decreases in comparison to a single YIG film due to the resistive losses attributed to the microwave eddy currents induced in a thin Pt layer by a propagating spin wave. By placing near the Pt layer of a highly conducting metal plate, we suppressed the excitation of the eddy currents in platinum and, thus, achieved a large increase (up to 1000 times) in the transmission of the Damon-Eshbach surface spin wave. 

\begin{figure}[t]
\includegraphics[width=8.5 cm]{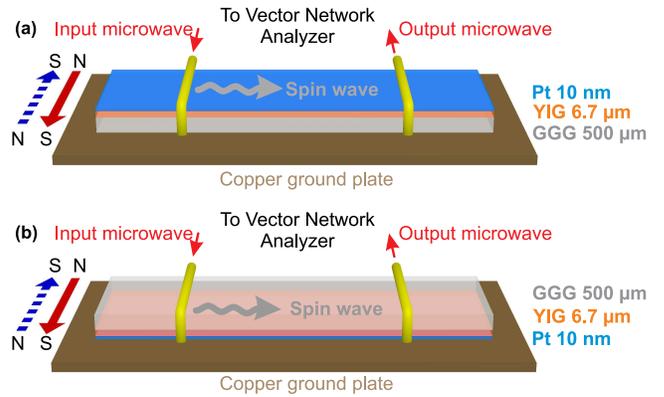}
\caption{\label{fig_setup} Sketch of the experimental setup. (a) The conventional SW excitation geometry -- a YIG-Pt bi-layer is placed near the  microstrip antennas. (b) The inverted SW excitation geometry -- the YIG-Pt bi-layer is close to the ground plate of the microstrip structure.}
\end{figure}

The experiments were performed using 2\,mm wide and 15\,mm long SW waveguides prepared from a 6.7\,$\mu$m thick YIG film grown by liquid phase epitaxy on a 500\,$\mu$m thick gallium-gadolinium garnet (GGG) substrate. A Pt layer of 10\,nm thickness was deposited on top of the YIG surface. Spin waves were emitted and received by two short-circuited gold-wire antennas of 50\,$\mu$m diameter placed 7.6\,mm apart from each other (see Fig.~\ref{fig_setup}). Spin-wave propagation was studied in a frequency band from 6.2 to 6.5\,GHz using a vector network analyzer (Anritsu MS46322A). A relatively small level of the applied microwave power of 0.5\,mW was chosen to prevent the development of non-linear effects for the propagating spin waves. An external bias magnetic field $H_0=1600$\,Oe was applied in the YIG film plane perpendicularly to the SW propagation direction. Thus, the geometry of the Damon-Eshbach (DE) surface spin wave was implemented.\cite{Serga2010} 

\begin{figure}[t]
\includegraphics[width=8.5 cm]{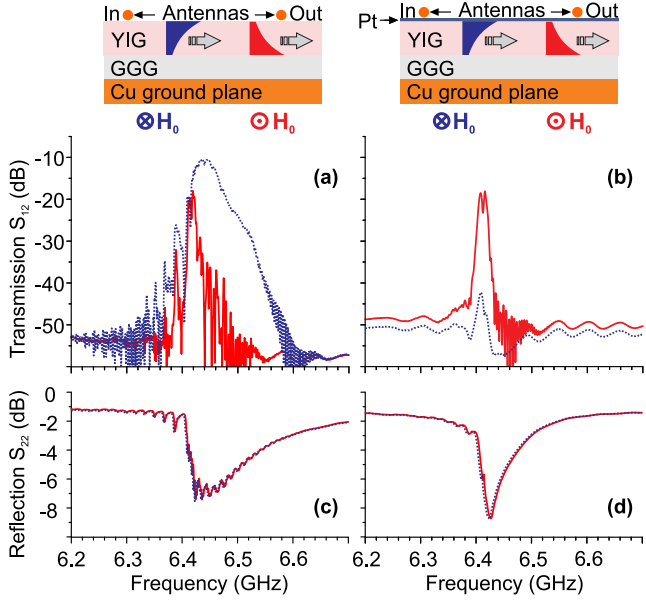}
\caption{\label{fig2} Conventional excitation geometry. A YIG film is positioned directly under the antennas. The localization of a Damon-Eshbach wave depends on the direction of the bias magnetic field $H_0$ (see insets). Dotted/blue lines correspond to the bias magnetic field pointing into the drawing plane. Solid/red lines correspond to the bias magnetic field pointing out of the drawing plane. (a) Transmission ($S_{12}$) characteristic of the bare YIG waveguide. (b) Transmission ($S_{12}$) characteristic of the YIG waveguide covered with 10\,nm Pt layer. (c) Excitation efficiency by the input antenna ($S_{22}$) for the bare YIG waveguide. (d) Excitation efficiency by the input antenna ($S_{22}$) for the YIG waveguide covered with 10\,nm Pt layer. \looseness=-1}
\end{figure}

The reference experiment was performed in a conventional excitation geometry (as in Ref.~\onlinecite{Wang2011}), when a YIG film lays directly under the antennas (see Fig.~\ref{fig_setup}(a) and insets in Fig.~\ref{fig2}). In the case of a bare YIG film -- i.e. without any Pt coating (Fig.~\ref{fig2}(a)), the DE spin wave, which propagates near the YIG surface touching the antennas (see the sketch of spin-wave distribution over the YIG film thickness in inset in Fig.~\ref{fig2}(a)), possesses relatively low minimal transmission losses of around 10\,dB at 6.42\,GHz. The DE wave localized at the opposite film surface shows one order of magnitude weaker transmission. This well known effect relates to the nonreciprocal excitation of a DE wave, whose efficiency depends on the relative orientation of the bias magnetic field $H_0$ and the SW propagation direction.\cite{Serga2010, Schneider2008, Demidov2009} In spite of the pronounced difference in the transmission characteristics, the efficiencies of SW excitation, which are inversely proportional to the microwave reflection from the input antenna $S_{22}$, are identical for two opposite orientations of the bias magnetic field (see Fig.~\ref{fig2}(c)). That is because in both cases the input antenna radiates spin waves in two opposite directions but only one of the emitted waves reaches the output antenna. 

In the case of the Pt-covered YIG film, the excitation efficiency remains almost the same as for the bare YIG waveguide (compare Fig.~\ref{fig2}(d) to Fig.~\ref{fig2}(c)) but the transmission of the DE wave localized near the Pt layer is strongly suppressed (see the dashed curve in Fig.~\ref{fig2}(b)). This suppression can be associated either with eddy currents excited in the high-resistive ($R_{\text {Pt}}=218$\,Ohm) Pt layer by stray fields of the propagating spin wave or with the spin-pumping effect at the YIG-Pt interface.

\begin{figure}[t]
\includegraphics[width=8.5 cm]{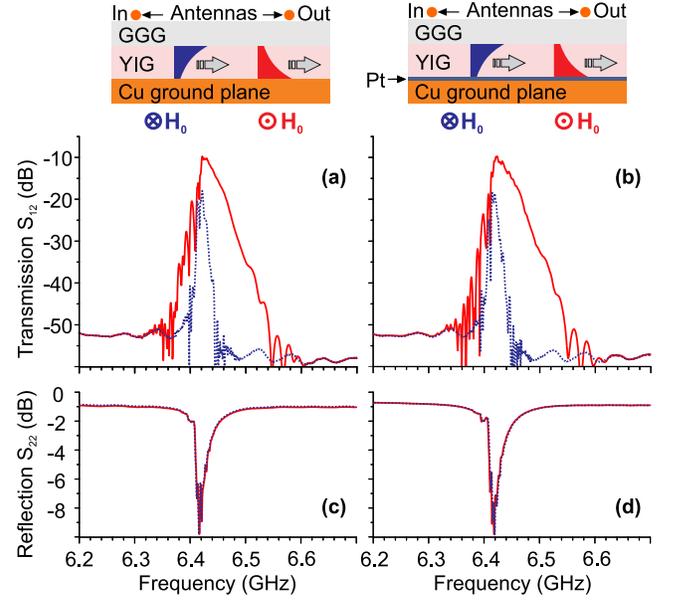}
\caption{\label{fig3} Inverted excitation geometry. The YIG waveguide is positioned directly on the copper ground plate. Dotted/blue lines correspond to the bias magnetic field pointing into the drawing plane. Solid/red lines correspond to the bias magnetic field pointing out of the drawing plane. (a) Transmission ($S_{12}$) characteristics of the bare YIG waveguide. (b) Transmission ($S_{12}$) characteristics of the YIG waveguide covered with 10\,nm Pt layer. (c) Excitation efficiency of the input antenna ($S_{22}$) for the bare YIG waveguide. (d) Excitation efficiency of the input antenna ($S_{22}$) for the YIG waveguide covered with a 10\,nm Pt layer. Insets demonstrate schematic representations of the surface spin wave localizations for different directions of a bias magnetic field.
}
\end{figure}

At the second stage of our studies, in order to distinguish between these two damping mechanisms, the YIG film was facing the ground plate (see Fig.~\ref{fig_setup}(b) and Fig.~\ref{fig3}). It is expected that if the Pt film is placed in direct contact with the bulk copper plate, than the eddy currents will mostly flow in the low resistive ground plate. Hence, no additional Ohmic losses due to currents in Pt are expected. At the same time, the spin pumping effect, which can potentially contribute to the excessive SW damping, should remain unchanged. The SW excitation efficiencies shown in Fig.~\ref{fig3}(c) and Fig.~\ref{fig3}(d) for the bare and Pt-covered YIG waveguides are close to those obtained in the conventional excitation geometry (compare with Figs.~\ref{fig2}(c) and \ref{fig2}(d)). At the same time, one can see from Figures~\ref{fig3}(a) and \ref{fig3}(b) that the wave, which propagates close to the Pt layer, experiences now no excessive  damping and shows the same transmission losses as the wave in the bare YIG film. 
This fact explicitly confirms our assumption about the significant contribution of the SW induced eddy currents to the SW damping in Pt-YIG bi-layers. \looseness=-1

It should be noted that in the proposed excitation geometry spin waves are excited by microwave currents flowing in the ground plate rather than through the microstrip antenna, which is separated from the YIG film by the thick GGG layer. It decreases the excitation efficiency of short-wavelength spin waves because the microwave currents are localized more weakly in the extended ground plate than in the narrow microstrip. However, the change in the electrodynamic boundary conditions due to placing a highly conducting layer near the surface of the YIG film \cite{Parekh1985} makes the DE dispersion relation steeper, i.e., increases the eigen-frequencies of the excited long-wavelength spin waves. Moreover, the related increase in the SW group velocity decreases the SW propagation time and consequently the SW transmission losses. As a result, the SW transmission frequency band remains rather wide. \looseness=-1

\begin{figure}[t]
\includegraphics[width=8.5 cm]{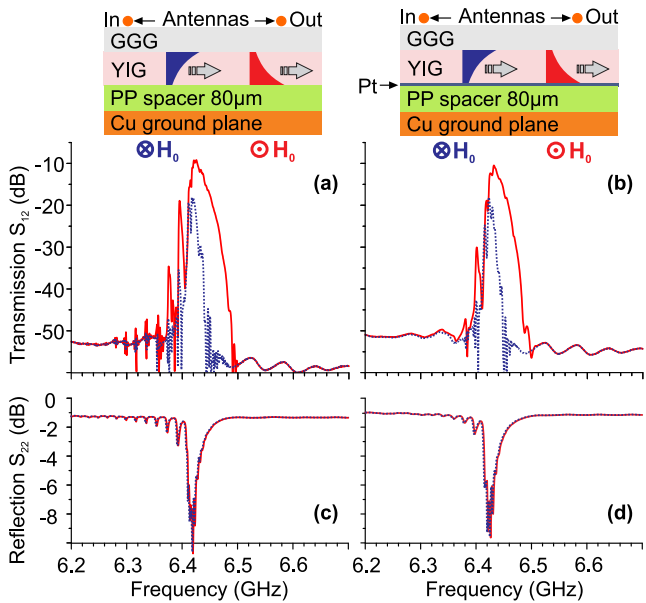}
\caption{\label{fig4} Inverted excitation geometry. The YIG waveguide is positioned near the copper ground plate. An additional 80\,$\mu$m thick dielectric polypropylen (PP) spacer layer is introduced to electrically decouple the Pt layer from ground. Dotted/blue lines correspond to the bias magnetic field pointing into the drawing plane. Solid/red lines correspond to the bias magnetic field pointing out of the drawing plane. (a) Transmission ($S_{12}$) characteristics of the bare YIG waveguide. (b) Transmission ($S_{12}$) characteristics of the YIG waveguide covered with 10\,nm Pt layer. (c) Excitation efficiency of the input antenna ($S_{22}$) for the bare YIG waveguide. (d) Excitation efficiency of the input antenna ($S_{22}$) for the YIG waveguide covered with a 10\,nm Pt layer. Insets demonstrate schematic representations of the surface spin wave localizations for different directions of a bias magnetic field.
}
\end{figure}

The described geometry enables the surface spin waves to propagate rather long distances through the Pt-covered YIG-film waveguides. However, no application of a dc electric current to such a structure is possible because of electric contact between the Pt layer and the Cu ground plate. In order to settle the issue, a 80\,$\mu$m thick polypropylene spacer was placed between the sample and the ground plate (see insets in Figs.~\ref{fig4}(a) and \ref{fig4}(b)). In this case, the minimal transmission losses (Figs.~\ref{fig4}(a) and \ref{fig4}(b)) remain the same as in the previous measurements (see Figs.~\ref{fig2}(a), \ref{fig3}(a) and \ref{fig3}(b)), but the transmission bandwidth turns to be significantly narrower. The reasons for this narrowing are obvious. A microwave magnetic field induced by microwave currents flowing in the ground plate is becoming more uniform with increase in the distance to the ground plate. This leads to a reduction of the excited SW wavenumbers and, alongside with the vanishing of the effect of metallization, to a decrease in the frequency bandwidth.
However, the coupling between the Pt layer and the ground plate appears to be high enough to shunt the Pt film and reduce, thus, the eddy currents related SW losses. 

At the final stage, we used the circuit shown in inset in Fig.~\ref{fig4}(b) in order to check the ability to control the SW damping by a dc electric current applied to the Pt layer. To avoid spurious heating effects the experiment was performed in the pulsed regime: overlapping input microwave pulse (20\,ns) and dc current pulse (350\,ns) were applied with 1\,kHz repetition rate. It is remarkable that in the studied structure the best SW transmission corresponds to the case when the DE wave is localized at the YIG-Pt interface. In the absence of the externally applied dc electric current, such localization results in rather high  dc voltages induced in the Pt layer by the combined action of the spin pumping and inverse spin Hall effects. However, no influence of both positive and negative electric currents $I_{\text {dc}} = \pm 370$\,mA on the SW amplitude was measured in the entire SW frequency band.

In conclusion, we have demonstrated that by using a novel spin-wave circuitry, where a Pt-covered YIG-film waveguide faces the ground plane of a microstrip line structure, it is possible to reduce the excitation of SW-related eddy currents in the Pt layer and achieve, thus, in a YIG-Pt bi-layer the same transmission losses for the Damon-Eshbach surface spin wave as in a bare YIG film. At the same time, no influence of a dc electric current applied to the Pt layer on the amplitude of SW pulses propagating in the relatively thick YIG waveguide of micrometer thickness was observed in our experiments.

Financial support by EU-FET InSpin 612759 and by the Australian Research Council is gratefully acknowledged.

\end{document}